\documentclass[acmlarge,manuscript,screen]{acmart}

\AtBeginDocument{%
  \providecommand\BibTeX{{%
    \normalfont B\kern-0.5em{\scshape i\kern-0.25em b}\kern-0.8em\TeX}}}

\setcopyright{acmlicensed}
\copyrightyear{2024}
\acmJournal{DGOV}
\acmYear{2024}
\acmDOI{XXXXXXX.XXXXXXX}

\usepackage{longtable}
\usepackage{tabularx}

\begin{document}

\title{Generative AI Literacy: Twelve Defining Competencies}
% \title{Twelve Defining Competencies of Generative AI Literacy}

\author{Ravinithesh Annapureddy}
\authornote{Both authors contributed equally to this research.
}
\email{ravinithesh.annapureddy@idiap.ch}
\orcid{0000-0003-4383-020X}
\affiliation{%
  \institution{Idiap Research Institute}
  \streetaddress{Rue Marconi 19}
  \city{Martigny}
  \country{Switzerland}}
\affiliation{%
  \institution{École Polytechnique Fédérale de Lausanne (EPFL)}
  \streetaddress{Rte Cantonale, INN 116, station 14}
  \city{Lausanne}
  \country{Switzerland}}
\email{ravinithesh.annapureddy@epfl.ch}
\author{Alessandro Fornaroli}
\orcid{0000-0002-4231-8231}
\authornotemark[1]
\email{alessandro.fornaroli@idiap.ch}
\affiliation{%
  \institution{Idiap Research Institute}
  \streetaddress{Rue Marconi 19}
  \city{Martigny}
  \country{Switzerland}
  \postcode{1920}
}

\author{Daniel Gatica-Perez}
\orcid{0000-0001-5488-2182}
\affiliation{%
  \institution{Idiap Research Institute}
  \streetaddress{Rue Marconi 19}
  \city{Martigny}
  \country{Switzerland}}
\affiliation{%
  \institution{École Polytechnique Fédérale de Lausanne (EPFL)}
  \streetaddress{Rte Cantonale, INN 138, station 14}
  \city{Lausanne}
  \country{Switzerland}}
\email{gatica@idiap.ch}


\begin{abstract}

This paper introduces a competency-based model for generative artificial intelligence (AI) literacy covering essential skills and knowledge areas necessary to interact with generative AI. The competencies range from foundational AI literacy to prompt engineering and programming skills, including ethical and legal considerations. These twelve competencies offer a framework for individuals, policymakers, government officials, and educators looking to navigate and take advantage of the potential of generative AI responsibly. Embedding these competencies into educational programs and professional training initiatives can equip individuals to become responsible and informed users and creators of generative AI. The competencies follow a logical progression and serve as a roadmap for individuals seeking to get familiar with generative AI and for researchers and policymakers to develop assessments, educational programs, guidelines, and regulations.

\end{abstract}




\received{28 January 2024}
\received[revised]{01 April 2024}
\received[accepted]{18 May 2024}

\maketitle

\section{Introduction}
With the rapid spread of Artificial Intelligence (AI) systems across all domains, the concept of \textit{AI literacy} (``a set of competencies that enables individuals to critically evaluate AI technologies''\cite{Long2020}) has become increasingly important and necessary in the past few years. The legislative work on the European AI Act \cite{EUAIACT2021} done by the European Parliament and European Commission has also contributed to increasing attention towards the risks and challenges posed by systems and tools based on AI models, as well as viable ways to regulate them \cite{Helberger2023, Heidelberger2023aiact}.

Generative models have found applications across many sectors, reflecting their adaptability and potential impact \cite{Sallam2023, Ray2023, Chungkwan2023}. As governments worldwide increasingly digitalize their operations and services, there is a growing intersection between AI and governance. Generative AI technologies can transform communication, public engagement, and decision-making processes with and within governmental bodies \cite{Bruce_2023}. Understanding the implications, challenges, and opportunities presented by generative AI is vital for researchers and practitioners in the field of digital government to make informed decisions about its usage and adoption. At the same time, a literate workforce is better equipped to identify and mitigate potential risks, ensuring that the deployment of generative AI in government processes is accompanied by risk assessment and mitigation strategies. With the transformation of the digital government arena, generative AI literacy is essential for future-proofing government operations. Professionals who understand generative AI can proactively adapt to emerging technologies, ensuring sustained relevance and effectiveness of digital government initiatives.

The concept of AI literacy has also emerged as a fundamental set of competencies to have during the AI era. However, the recent explosion of generative AI, spearheaded by ChatGPT, the Large Language Model (LLM) developed by OpenAI, has created the need for a more definite set of skills, abilities, and knowledge, specific to the scope and applications of generative AI.
The need for specific generative AI literacy arises from the following factors. First, the functionality to generate new text, images, music, videos, and code requires a distinct understanding of what generative AI can and cannot do compared to other existing AI tools. Second, generative AI's transdisciplinarity has implications for various fields, such as writing, design, and scientific research \cite{Ray2023}. Thus, an awareness of how these systems operate to produce output that meets certain criteria, coherence, and creativity is needed. In the third place, the outputs of generative AI raise critical questions on authorship, ownership, and originality. Thus, understanding generative AI involves not just the technical aspects of the algorithms or tool utilization but also bringing significant attention to the ethical and practical considerations in its use. Finally, generative AI tools are increasingly becoming accessible and user-friendly, necessitating additional aspects of AI literacy for the general public, not just for people specialized in technology.

The purpose of this paper is to propose a competency-based model that can delineate a general concept of \textit{generative AI literacy}, in which the model can be used as a framework to assess the ability of users to understand, interact with, and create generative AI models. This paper aims to fill a gap at the intersection of the topics of AI literacy and generative AI, as no set of competencies specific to the use of generative AI had been previously proposed. 

The twelve competencies identified in this paper can be intended as a complement to the existing framework on AI literacy, providing additional guidance on the specific toolbox that a user of generative AI tools needs. This concept of generative AI literacy is associated with that of \textit{prompt engineering}, which is the ability to design effective prompts for generative AI models to produce desired outputs \cite{Reynolds2021}.  However, while prompt engineering is one of the competencies and is crucial for generative AI literacy, the proposed model goes well beyond it and the more technical features of generative AI, and considers the contextual and situational aspects. This work can therefore serve as a starting point for future research on the applications of generative AI and its adoption in public government, institutions, and organizations. 

The rest of the paper is structured as follows: after explaining the methodology in Section \ref{sec:methodology}, Section \ref{sec:previous-research} gives an overview of existing literature concerning the concepts of AI literacy and generative AI, as well as laying the theoretical framework of a competency model. Section \ref{sec:model} proceeds to show an overview of the 12 competencies. Section \ref{sec:competencies} includes detailed descriptions of each competency, providing illustrative examples. Section \ref{sec:discussion} discusses the benefits and shortcomings of this proposed competency model, together with potential applications and ideas for future research. Section \ref{sec:conclusion} presents our conclusions.

\section{Methodology} \label{sec:methodology}

To begin this work, a preliminary database search was conducted in order to identify the state of the art on generative AI literacy. Five scientific databases, specifically \textit{ACM Digital Library}\footnote{\url{https://dl.acm.org/}}, \textit{IEEExplore}\footnote{\url{https://ieeexplore.ieee.org/}}, \textit{Clarivate Web of Science}\footnote{\url{https://www.webofscience.com/wos}}, \textit{Scopus}\footnote{\url{https://www.scopus.com/}}, and \textit{Dimensions}\footnote{\url{https://app.dimensions.ai/}}, were searched with the specific query 
\textit{``generative AI literacy'' OR ``generative artificial intelligence literacy''}. Given the low amount of results (6 records in total, as described in Section \ref{sec:database-search}), all the articles are described in Table \ref{tab:records}. The scarcity of results highlighted the novelty of the concept and the current gap in academic literature.

Consequently, given the novelty of the topic, a long-standing review approach on generative AI literacy is not feasible at this moment, and instead, a non-systematic, comprehensive literature review was carried out on studies written in English, including seminal papers and systematic reviews on the following topic areas:
\begin{itemize}
    \item AI literacy
    \item Systematic reviews on the applications of generative AI
    \item Prompt engineering
    \item Competency models
\end{itemize}

Following the database search and literature review, we identified a list of competencies that can help define the concept of \textit{generative AI literacy}, covering different aspects that emerged from the literature.
The findings are summarized in Section \ref{sec:model}, which also includes some speculative potential implications of having acquired (or not) the competencies. Given the novelty of the topic, a speculative approach was chosen for discussing potential implications, relying on informed opinions and current trends due to the scarcity of previous research on this specific issue.

Each of the competencies is then described in detail, and provided with an illustrative example. These descriptions are included in Section \ref{sec:competencies}.

\section{Literature Review} \label{sec:previous-research}
\subsection{Preliminary Database Search}\label{sec:database-search}

As explained in Section \ref{sec:methodology}, an initial database search was conducted in December 2023, using the specific strings ``generative AI literacy'' \textit{or} ``generative artificial intelligence literacy''. In particular, the following databases were searched, giving these results:
\begin{itemize}
    \item ACM Digital Library: 1 record found \cite{ElZanfaly2023}.
    \item IEEExplore: 2 records found \cite{Putjorn2023, Dadhich2023}.
    \item Clarivate Web of Science: 2 records found \cite{ElZanfaly2023, Bozkurt2023}.
    \item Scopus: 2 records found \cite{ElZanfaly2023, Putjorn2023}.
    \item Dimensions: 4 records found \cite{ElZanfaly2023, Noh2023, Putjorn2023, Zhai2023}.
\end{itemize}

After removing duplicates, this search resulted in a total of 6 records, which have been included and summarized in Table \ref{tab:records}. The table includes a short summary of the contents of each of the studies and a description of how the concept is defined.

\begin{table}
  \caption{Records found in the database search on ``generative AI Literacy''}
  \label{tab:records}
  \begin{tabularx}{\textwidth}{XXX}
    \toprule
Study and reference & Short summary 
& Definition of generative AI literacy \\
\midrule
Bozkurt, 2023 \cite{Bozkurt2023} & Systematic review and bibliometric analysis of AI research in education, highlighting recent trends and directions for research. & Concept introduced and loosely defined as ``proficiency in understanding, interacting with, and critically evaluating generative AI technologies'', which ``entails not only knowing how to use AI-driven tools but also understanding the ethical considerations, biases, and limitations inherent in such systems''\cite{Bozkurt2023}. \\ \\

Dadhich and Bhaumik, 2023 \cite{Dadhich2023} & Quantitative study linking generative AI literacy, algorithmic thinking, cognitive divide, and pedagogical knowledge in higher-education students. & Concept conflated with that of general AI literacy, and not clearly defined. \\ \\

Nyaaba and Zhai, 2023 \cite{Zhai2023} & Study on the outcomes of a professional development webinar on generative AI for teacher educators in Ghana. & Concept not clearly defined. \\ \\

Noh and Han, 2023 \cite{Noh2023} & Study on the implementation of a generative AI literacy education program for pre-service secondary teachers. & The main paper is in Korean language. In the English abstract, the concept is not clearly defined. \\ \\

Putjorn and Putjorn, 2023 \cite{Putjorn2023} & Study on the perception of young, teenage learners of generative AI, with workshops and questionnaires. & Concept conflated with that of general AI literacy. \\ \\

El-Zanfaly, Huang, and Dong, 2023 \cite{ElZanfaly2023} & Study introducing a tangible interface for human-AI co-creation using sand as a medium, aiming to enhance understanding of generative AI applications in design. & Concept introduced but not defined. \\ \bottomrule
\end{tabularx}
\end{table}

As observed from Table \ref{tab:records}, there is not a well-established concept of \textit{generative AI literacy} in the current academic literature. The few existing papers tend to conflate it with the more general and better-established concept of \textit{AI literacy}.

\subsection{AI literacy and generative AI}

Several studies in recent years have explored the concept of AI Literacy \cite{Ng2021, yi2021}, and a number of systematic reviews have been published on the topic \cite{Long2020, Ng2021, Casal-Otero2023}.
An often-cited definition of \textit{AI literacy} is given by Long and Magerko \cite{Long2020} as ``a set of competencies that enables individuals to critically evaluate AI technologies; communicate and collaborate effectively with AI; and use AI as a tool online, at home, and in the workplace'' \cite{Long2020}. The authors also provide a list of 16 core competencies that form the basis of AI literacy, as well as a set of design considerations that should help design readily accessible AI tools. 

Nonetheless, as also indicated by the preliminary database search, comparatively few studies explore the concept of AI literacy in the specific context of generative AI \cite{kreinsen_schulz_2023}. This is particularly relevant as, on the other hand, there has been a surge in the scholarly literature concerning generative AI, i.e., the deep-learning models aimed at generating original content that aims at simulating human-made content. Specifically, these models include Generative Adversarial Networks (GANs) \cite{Creswell2018}, Variational Autoencoders (VAEs) \cite{kingma2019}, and autoregressive models, such as transformers \cite{wolf2020}, along with others.
In particular, the emergence of ChatGPT, together with its capabilities and wide range of applications, has led to great media and scientific attention towards this model and LLMs in general \cite{Ray2023, Wu2023}.
Multiple systematic reviews have been published on the topic \cite{Ray2023, Sallam2023}, analyzing and discussing the impacts of ChatGPT and generative AI in general on different areas including education \cite{Kasneci2023, Sallam2023, Preiksaitis2023, BaidooAnu2023, Chungkwan2023, Sok2023, Kohnke2023}, healthcare \cite{Aydin2022, Thirunavukarasu2023, Vaishya2023, Li2023Healthcare, Liu2023Clinical}, research \cite{Sok2023, Rahman2023,Ruksakulpiwat2023}, and society \cite{Baldassarre2023}, among others. 

Hence, the academic interest in ChatGPT, LLMs, and generative AI, in general, has been soaring, but at the same time, more research questions arise and need to be addressed \cite{VanDis2023, Dwivedietal2023}. Despite this, the concept of AI literacy has not evolved, and existing frameworks do not differentiate between generative and predictive models \cite{Long2020, long2023large}.

Table \ref{tab:literature} summarizes some studies relevant to the definition of the concept of \textit{generative AI literacy}, stemming from the main topics of AI literacy, prompt engineering, and applications of generative AI models. 

\begin{longtable}{p{0.2\linewidth}p{0.4\linewidth}p{0.125\linewidth}p{0.175\linewidth}}
\caption{Studies relevant to the definition of a concept of generative AI literacy}\\

\toprule
    Study and reference & Short summary & Topic area & Type of Study \\ \midrule
\endhead  

Long and Magerko, 2020 \cite{Long2020} & Systematic review and framework for AI literacy, outlining core competencies and design considerations.
& AI literacy & Systematic review and conceptual paper \\ \\

Ng et al., 2021 \cite{Ng2021} & Systematic review on AI literacy. & AI literacy & Systematic review \\ \\

Casal-Otero et al., 2023 \cite{Casal-Otero2023} & Systematic review on the integration of AI in K-12 education worldwide. & AI literacy & Systematic review \\ \\

Schüller et al., 2023 \cite{schuller2023} & Framework for a taxonomy including both data and AI literacy, highlighting their role in informed decision-making. & AI literacy & Conceptual paper \\ \\

Yi, 2021 \cite{yi2021} & Framework for type, competence, and purpose of AI literacy. & AI Literacy & Conceptual paper \\ \\

Kreinsen et al., 2023 \cite {kreinsen_schulz_2023} & Discussion of the similarities, intersections, implications, and applications of Digital Literacy, Data Literacy, and AI Literacy in teacher education. & AI literacy & Conceptual Paper \\ \\ 

Giray, 2023 \cite{Giray2023} & Study delineating some principles and techniques of prompt engineering for ChatGPT. & Prompt \quad\quad engineering & Conceptual study and guidelines \\ \\

Wang et al., 2023 \cite{Wang2023} & Systematic review on visual prompt engineering and large vision models. & Prompt \quad\quad engineering & Systematic review \\ \\

Hwang, Lee, and Shin, 2023 \cite{hwang2023} & Study introducing the concept of ``prompt literacy'', related to the skill of prompt engineering. & Prompt \quad\quad engineering & Experimental study \\ \\

Lo, 2023 \cite{Lo2023} & Guide on prompt engineering, including discussion of advantages and risks. & Prompt \quad\quad engineering & Conceptual study \\ \\

Ray, 2023 \cite{Ray2023} & Systematic review of studies on ChatGPT under different aspects & Studies on \quad generative AI & Systematic review \\ \\

Wu, 2023 \cite{Wu2023} & Study on the history and development of ChatGPT and its applications & Studies on \quad generative AI & Descriptive paper \\ \bottomrule
    \label{tab:literature}
\end{longtable}

\subsection{Competency acquisition}

The concept of competency has been a subject of extensive research in academic literature, with various definitions proposed over time \cite{Delamare2005, cernucsca2007, Woodruffe1993, Hoffmann1999}. Succar et al. \cite{Succar2013} proposed a Triple-A competency model that emphasizes the importance of acquiring, applying, and assessing competencies. Possessing competencies often carries positive implications and can, in principle, be assessed \cite{Shavelson2010}. From an organizational perspective, competencies can be organized into a \textit{competency model}, which Chouhan and Srivastava \cite{Chouhan2014} define as ``a valid, observable, and measurable list of knowledge, skills, and attributes that are demonstrated through behavior to achieve exceptional performance in a specific work setting.''

Competencies are conceptualized to be applied in practice, being useful in specific contexts \cite{Woodruffe1993}. It is also important to take into account that acquiring a competency is a process, which is not unique given the huge diversity of contexts in which users are acquiring it \cite{Batalden2002, Hoffmann1999}. In this sense, this paper adopts the definition given by the National Institute of Health (NIH), according to which ``Competencies are:
Knowledge (Information developed or learned through experience, study, or investigation),
Skills
 (The result of repeatedly applying knowledge or ability); 
Abilities
(An innate potential to perform mental and physical actions or tasks);
Behaviors
(The observable reaction of an individual to a certain situation)'' \cite{NIH2020}.

\subsection{Need for a Generative AI Literacy competency model}

While previous studies have explored AI literacy in various contexts and investigated the capabilities and limitations of generative AI models, the database search showed the absence of literature defining the concept of \textit{generative AI literacy} and also a comprehensive framework that systematically addresses competency acquisition in navigating and utilizing generative AI tools. 

At the same time, the existing frameworks on AI literacy tend to be quite generic, failing to address the specificities of generative AI tools. Also, while both generative models, which have the capacity to synthesize new content, and predictive models, which extrapolate future outcomes based on previous patterns, are trained on vast datasets, the existing literacy frameworks do not differentiate between them. From the technical point of view, one needs to be aware of the probabilistic mechanisms underpinning the synthesis tools in generative models. On the other hand, predictive models necessitate an understanding of statistical modeling and forecasting techniques to distinguish the mode of operation of AI systems they encounter in practice. Similarly, ethical considerations further underscore this need for distinction. Generative models are ingrained with the potential to fabricate highly realistic content and give rise to ethical dilemmas about misinformation, authenticity, and societal impact. Thus, in addition to the knowledge of the ethical issues from the technical side, one has to build an understanding of the origins and mechanisms of AI-generated content to mitigate potential harm on the issues inherent to the dissemination and consumption of generated content. Lastly, incorporating differentiation between generative and predictive models into professional training programs and educational initiatives is crucial to address the challenges of each model type and be better prepared for using generative AI in typical applications, or in specific careers in AI-related fields.

The rise of unique needs pertaining to the domain of generative AI has motivated us to determine precise literacy competencies that individuals should develop in order to master generative AI tools.

\section{A Competency-Based Model for Generative AI Literacy} \label{sec:model}

\begin{figure}[ht]
    \centering
    \includegraphics[width=0.8\textwidth]{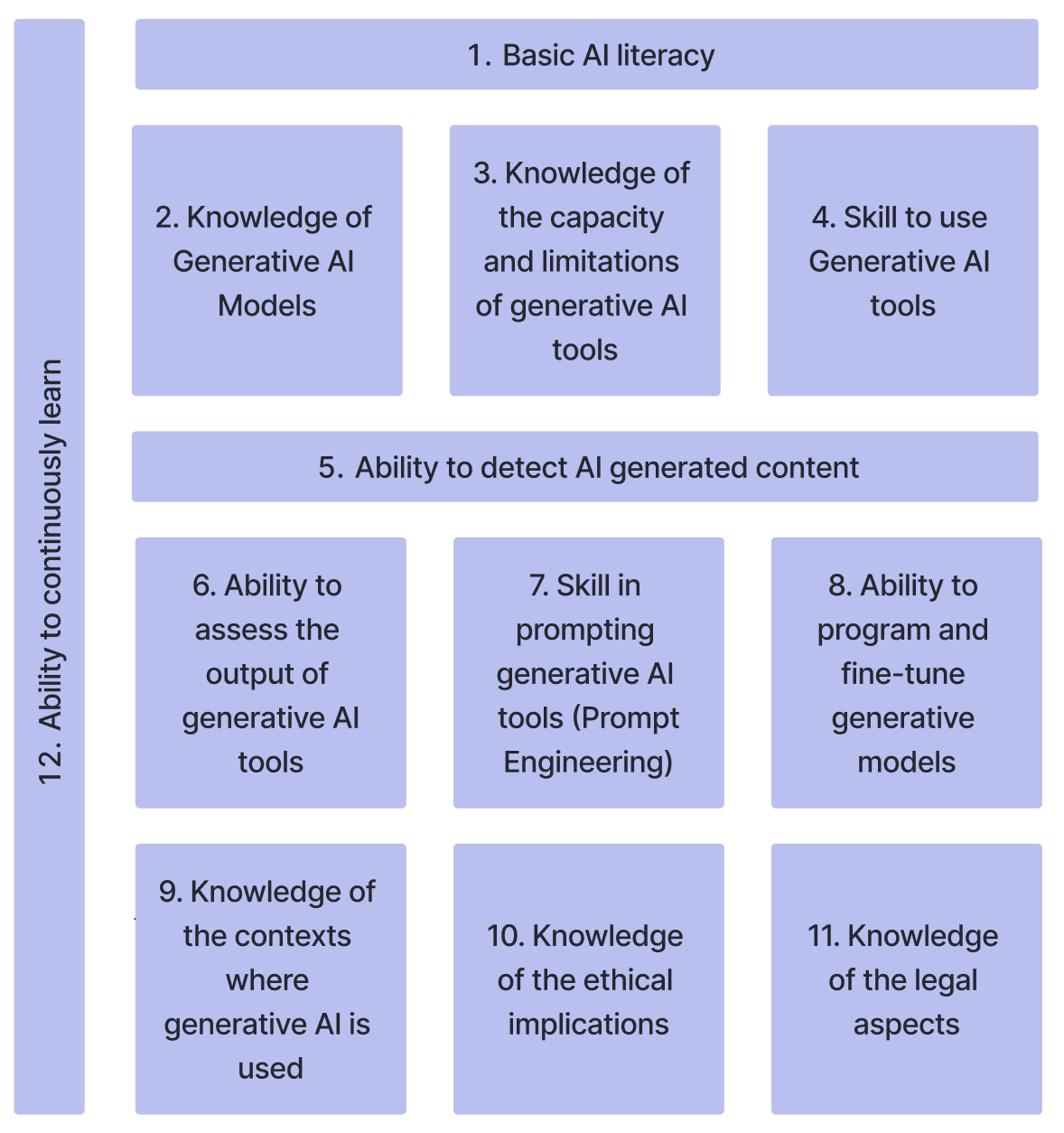}
    \caption{Diagram of the 12 generative-AI competencies}
    \Description{A Diagram showing the 12 generative-AI competencies}
    \label{fig:diagram}
\end{figure}

As previously mentioned, the rapid rise of generative AI tools and their widespread adoption in various disciplines have created a unique set of skills specific to these novel tools, which go beyond the traditional competencies associated with data literacy, digital literacy, or AI literacy. This paper aims to identify a set of competencies that are necessary for users to develop to effectively comprehend and utilize generative tools. This paper adopts the aforementioned definition of \textit{competency} given by the National Institute of Health, which groups competencies into knowledge, skills, abilities, or behaviors \cite{NIH2020}. Alongside, this paper defines generative AI literacy as a long-term, continuous process by which several core competencies, allowing to effectively and responsibly understand, assess, and work with generative AI technologies, have been identified, acquired, practiced, and mastered.

Based on existing literature on AI literacy, digital literacy, prompt engineering, generative AI, and related topics, including surveys and reviews, peer-reviewed publications, gray literature, and books, this work identifies twelve specific competencies to define the concept of \textit{generative AI literacy} acknowledging the varied nature of engaging with generative AI.

As illustrated by Figure \ref{fig:diagram}, the ordering of these competencies is purposeful and follows a logical progression to represent a structured learning path reflecting a continuum of learning, as continuous self-learning is transversal to all the other capabilities. The initial focus is on 
creating the general basis of knowledge and skill to know and use generative AI tools, followed by knowing how to detect AI-generated content. These develop into more technical skills related to the use of generative AI, with content assessment, prompt engineering, and programming. Then, the three interrelated competencies regarding the knowledge of the contexts, ethical implication, and legal aspects, are intentionally placed towards the end to underline their significance. 
This approach allows individuals to progressively acquire skills and knowledge, encouraging a user-centric learning experience.

All the competencies of this Competency-Based Model for generative AI literacy are outlined in Table \ref{tab:genai-competencies}, with a more detailed explanation in Section \ref{sec:competencies}. 
Table \ref{tab:genai-competencies} includes examples of the positive implications of being proficient with each capability, as well as the negative implications of not acquiring the specific competency. As mentioned in Section \ref{sec:methodology}, these potential implications are mostly speculative, because of the novelty of the concepts presented and the unavailability of previous research. The \textit{positive implications of mastering the competencies} represent potential benefits of being skilled in each of the 12 competencies; conversely, the \textit{negative implications of not acquiring a competency} represent potential risks that may become real when an individual is not proficient in a given competency.

\begin{longtable}{p{0.02\linewidth}p{0.2\linewidth}p{0.22\linewidth}p{0.22\linewidth}p{0.22\linewidth}}
\caption{Competencies for Generative AI Literacy and their implications}\\

\toprule
    No. & Competency & Description & Positive Implications & Negative Implications \\ \midrule
\endhead
    1. & Basic AI literacy & Establishes a baseline understanding of AI concepts, laying down the groundwork for familiarity with generative AI. & Allows to engage in informed discussions about AI technologies and understand their impact. & Leads to misconceptions and uninformed decisions, hindering effective participation in AI-driven environments. \\ \\
    
    2. & Knowledge of generative AI models & Provides a peripheral understanding of the workings of generative AI. & Comprehend that generative AI creates content, distinguishing it from search engines that retrieve existing information. & Misconceptions and false beliefs about how generative AI operates and how its responses are generated. \\ \\
    
    3. & Knowledge of the capacity and limitations of generative AI tools & Equips individuals with the proficiency to assess the capabilities and constraints of generative AI tools. & Sets realistic expectations, leading to more effective and targeted application of generative AI tools. & Over-reliance or misjudging, potentially leading to misuse or missed opportunities. \\ \\
    
    4. & Skill to use generative AI tools & Promotes practical proficiency to effectively leverage generative AI tools in diverse contexts. & Allows to utilize the power of generative AI for creative tasks, problem-solving, and productivity. & Annoyance and underutilization of available tools. \\ \\
    
    5. & Ability to detect AI-generated content & Teaches the skill of discerning AI-generated content. & Verify content authenticity, maintaining trust in information sources. & May lead to misinformation, eroding trust and credibility. \\ \\
    
    6. & Ability to assess the output of generative AI tools & Provides the ability to critically assess output quality, relevance, and potential biases. & Verify whether the generated content is relevant and useful for their purposes & May lead to vulnerabilities by using irrelevant, inaccurate, or unhelpful content. \\ \\
    
    7. & Skill in prompting generative AI tools (Prompt Engineering) & Nurtures the creative aspect of working with generative AI, allowing individuals to tailor personalized outputs to specific objectives or creative efforts. & Allows to tailor AI outputs to specific needs and objectives. & May result in undesired outputs, limiting the practical value of generative AI tools. \\ \\
    
    8. & Ability to program and fine-tune & Provides the technical know-how necessary for the customization and optimization of generative AI models to suit specific needs. & Allows customization of AI models, adapting them to specific requirements and to achieve optimal performance. & May limit users to off-the-shelf solutions, hindering adaptability and innovation. \\ \\
    
    9. & Knowledge of the contexts where generative AI is used & Understand the diverse applications and limitations of generative AI across situations, institutions, and professions to assess the appropriateness of using generative AI tools. & Facilitates appropriate application of generative AI. & May lead to inappropriate use, potentially causing unintended consequences. \\ \\
    
    10. & Knowledge of the Ethical Implications & Augments technical proficiency with ethical considerations and inculcates a sense of responsibility by making individuals aware of the ethical considerations tied to the use of generative AI. & Promotes responsible use of generative AI, considering potential ethical concerns and societal impacts. & May lead to unintended harm, public backlash, or legal repercussions. \\ \\
    
    11. & Knowledge of Legal Aspects & Addresses legal dimensions, ensuring individuals operate within the bounds of intellectual property and other legal frameworks associated with the use of generative AI. & Ensures regulatory compliance, reducing the risk of legal issues and aiding in the responsible use of generative AI. & May result in legal consequences, damage to reputation, and barriers to innovation. \\ \\
    
    12. & Ability to continuously learn & Promotes a mindset of continuous learning to stay updated with evolving generative AI technologies, methodologies, and ethical considerations. & Fosters adaptability and keeps individuals up-to-date on dynamically evolving AI technologies. & May lead to obsolescence, limiting one's ability to harness the full potential of generative AI advancements. \\ \bottomrule
    \label{tab:genai-competencies}
\end{longtable}

\section{Generative AI Competencies}\label{sec:competencies}

This section describes each of the twelve competencies defined in the previous section in detail and provides an example of what an individual can do by acquiring that competency.

\subsection{Basic AI Literacy}
First and foremost, any framework for generative AI literacy should comprise more general AI literacy competencies and skills. Indeed, the knowledge and understanding of generative tools is in large part conditional to a familiarity with AI systems. Similarly, to assess a user's ability to understand, use, and evaluate generative tools, it is essential to consider their expertise regarding the broader area of AI software.

As noted, AI literacy does not equate to computing literacy, nor is computing literacy necessary for AI literacy, as individuals may recognize, learn about, and use AI without necessarily being able to program or having experience in computer science and engineering \cite{Long2020}.

\textbf{Example:} Individuals will be able to recognize different types of AI. This will enable them to evaluate the potential benefits and risks, understanding how AI can impact them or their business processes.

\subsection{Knowledge of generative AI models}

The first competency relating to generative AI pertains to specific knowledge of what it is. Growing media attention on generative models, specifically LLMs, has raised general awareness of AI. However, it has also led to increased confusion surrounding AI's true nature \cite{li2023chatgpt}.

Within the scope of the competency-based model for generative AI literacy, knowledge, and understanding of this concept can be defined as a competency with different levels of depth. It is primarily relevant to assess whether an individual has heard of generative AI or some examples of it, such as ChatGPT. Then, individuals should know how it works at a high level, in particular, that it is a statistical model, that it is trained on big data, and that it can generate different types of output. Users should also understand the differences between the user interface and the model itself, and that it is fundamentally different from a search engine, as it generates original content rather than retrieving existing one.

\textbf{Example:} Despite the extensive media coverage, ChatGPT lacks widespread recognition among diverse demographics, particularly among certain age groups and professions. Even among those who have heard of it, usage remains limited \cite{Watnick2023}.

\subsection{Knowledge of the capacity and limitations of generative AI tools}

Besides knowing what generative AI models \textit{are}, users should also understand what they \textit{can do}, including their potential and limitations. 
The increasing media presence of generative AI and AI-generated content has started to raise concern regarding the current and future capabilities of AI, increasing the worry that AI systems can be dangerous and used against the public good \cite{Chan2023-ud, Chan2023-bh}. It is, therefore, relevant that users are aware of the actual limitations as well as capabilities of generative AI, with their advantages, strengths, weaknesses, and potential harms. Moreover, it is beneficial to be able to recognize and understand patterns in generative AI and AI-generated content, identifying the main attributes of the models. 

Among the challenges posed by generative AI, users should be aware of privacy and security concerns, especially concerning personal information fed as input into the existing services, and impersonation and identity theft that can be facilitated by deep fakes. Another threat is posed by the potential misuse of technology as a technique to deceive people, including the fabrication of videos, audio, images, and text, for purposes that can span from phishing attacks to social engineering.

\textbf{Example:} The output of ChatGPT and other LLMs can often contain a mixture of facts and completely false and fabricated statements, and awareness of this is fundamental \cite{Alkaissi2023, Azamfirei2023}.

\subsection{Skill to use generative AI tools}

In terms of the competency-based model for generative AI literacy, a valuable skill for users is to learn how different generative tools work, and how to interact with new emerging generative models. This includes expertise in a specific model, the knowledge of different existing models tailored to specific needs, as well as the ability to easily learn and use new tools depending on need. 

The plethora of existing tools based on generative AI models has brought the need to be aware of the diversity of the existing generative tools and what they can do, as well as the modality and specificities of each one of them. Hence, a very valuable competency is to know which tool to use for a specific application, and how to use it to generate new content. This skill can be defined as somewhat technical, as it implies practicing the use of models, and can be easily improved with courses targeted to specific toolsets. Indeed, this has become a noteworthy skill to have in the labor market, as increasingly more companies require employees who can use and master generative AI \cite{eloundou2023, Budhwar2023}.

\textbf{Example:} A user who needs to generate images should start to learn and practice several tools for this, such as Midjourney\footnote{\url{https://www.midjourney.com/}}, Adobe Firefly\footnote{\url{https://www.adobe.com/products/firefly.html}} (available in Photoshop), DreamStudio\footnote{\url{https://dreamstudio.ai/}}, and others.

\subsection{Ability to detect AI-generated content}

With the advance of generative AI, it has become increasingly difficult to discern which contents have been entirely human-generated, AI-generated, or AI-enhanced \cite{Dwivedietal2023}. 
Hence, this competency concerns both being able to tell apart human-made from AI-made content, and knowing how to use AI detection software. Guidelines and techniques to help users identify which images, videos, and text have been created by AI \cite{Harding2023} are already being discussed. However, human detection of AI-generated photos and texts becomes increasingly difficult \cite{Nightingale2022}, making AI detectors always more significant. 
In recent years, there has been a surge in software and AI models to detect AI-generated content, both text (such as \textit{ZeroGPT}\footnote{\url{https://www.zerogpt.com/}}, \textit{GPTZero}\footnote{\url{https://gptzero.me/}}, \textit{PlagiarismCheck}\footnote{\url{https://plagiarismcheck.org/}}, \textit{GPT Radar}\footnote{\url{https://gptradar.com/}}, \textit{Writer.com}\footnote{\url{https://writer.com/ai-content-detector/}}, among others) \cite{Weber2023, Chaka2023,FuiHonnNah2023}, images (\textit{AI or Not}\footnote{\url{https://www.aiornot.com/}} or \textit{Is it AI?}\footnote{\url{https://isitai.com/ai-image-detector/}}) and videos \cite{Shehzeen2021, Qi2020}. 

At the same time, it is noteworthy that this task is becoming increasingly difficult, and individuals should ideally be aware of state-of-the-art tools to try to optimize their probability to correctly distinguish AI-generated from human content \cite{Gao2022, Baraheem2023, Groh2024}. In this sense, some of these detectors tend to overestimate the probability that a text, image, or video is AI-generated, while others underestimate it, and there is often a great divergence in results across different AI detection tools \cite{Weber2023}. Hence, users should be aware of the biases of the detectors they use, and possibly use a combination of them to verify the content.

\textbf{Example:} When seeing a video that looks questionable, people should be able to verify, through AI detectors and official sources, whether the video is authentic or fabricated. A recent example from 2023 is the case of a viral AI-generated video of the Pentagon on fire, which created some panic \cite{Collins2023, Haddad2023}.

\subsection{Ability to assess the output of generative AI tools}

After the users manage to generate output with generative tools, they should also be able to carry out a thorough assessment of it, in order to use it in a way that corresponds to their needs and expectations. This includes a verification of the content of the output with the aid of trusted sources, so to prevent or minimize the effect of model hallucinations \cite{Alkaissi2023, Athaluri2023, Azamfirei2023}. Critical thinking is paramount in this task, as users should be able to only select the content that is the most useful to them, discerning what information and format they should use, and how to adapt it for their desired applications. 

This competency is related to prompt engineering (the skill to effectively prompt generative AI models) and can be seen as being somewhat propaedeutic to it. Indeed, most prompt engineering techniques include an assessment of the output, but this does not necessarily imply that the user is fully proficient in prompt engineering.

\textbf{Example:} Individuals using LLMs to write essays on specific topics should check the facts presented in the output, to verify that they are not hallucinations of the model, and present these facts in a way that corresponds to their needs. 

\subsection{Skill in prompting generative AI tools (Prompt Engineering)}

One of the main competencies specific to the use of generative AI is that of prompt engineering, specifically a set of techniques and methods to write prompts as inputs of text-based generative models \cite{Reynolds2021}. Different sets of guidelines have been written on prompt engineering for ChatGPT and LLMs, highlighting its importance \cite{arora2022ask, Giray2023, si2023prompting, White2023}.
It is important to note that prompt engineering is not limited to LLMs, but to all types of generative AI that use text as input, including text-to-image models, as evidenced by the existing literature and guidelines for text-to-image prompt engineering \cite{Liu2022, Zhou2022-ic}. Prompt engineering consists of a broad skill set, with different methods that allow to effectively interact with generative models \cite{Saravia_2022}. A number of techniques have emerged for this purpose, including automatically generated prompts \cite{shin-etal-2020}, Chain of Thoughts prompting \cite{Wei2022, zhang2023multimodal, Kojima2022} and Tree of Thoughts prompting \cite{yao2023tree, long2023large}. Multimodal prompting, including images and other formats as input, has also emerged and increased in popularity \cite{Qiao2022prompting}

While it is beyond the scope of this paper to delve into the details of prompt engineering, given the abundance of academic writings and systematic reviews of the topic \cite{Chen2023, Gu2023, Liu2023}, being aware of different techniques to optimally prompt AI systems can allow users to be more efficient in their use of generative AI.

Most generative models that are deployed at the time, whether they are LLMs, text-to-speech, or text-to-image, rely on certain hyperparameters that can be adjusted by the users. In the case of ChatGPT, for example, these include the \textit{temperature}, which determines the “creativity” of the model, the maximum number of tokens to be used, and the \textit{top\_k} and \textit{top\_p} to limit the number of next word choices of the model and the probability of the next word choices, respectively \cite{Sarrion2023, Lo2023CLEAR}. Hence, fixing these parameters is an additional and more advanced skill to know when mastering prompt engineering. 

\textbf{Example:} In the context of image generation, the utilization of descriptive language, understanding the trade-offs between creativity and specificity, the possibility of segmenting longer prompts into smaller units, or the incorporation of negative words, can all lead to outputs that are different from the conventional results \cite{Pykes2023}.

\subsection{Ability to program and fine-tune}

The most high-level technical competency consists of more programming-related skills with respect to the development of generative AI models. These include the design of new models and architectures, proficiency in the use of common libraries for deep learning, and being able to adapt existing generative models and fine-tuning them to specific applications. People competent with this ability should be able to select appropriate models and adapt their architecture to the circumstances, select and prepare the training data, and know how to implement, train, test, and deploy a model. Most users will neither have nor need this ability for their specific goals, and still be literate on generative AI without it.  

\textbf{Example:} This ability is valuable for developing generative AI in specialized domains such as medical writing or dealing with languages other than English. Developers can adapt these models to equip them with a deeper understanding of domain-specific vocabulary and intricacies, leading to more impactful and tailored content generation or for particular linguistic nuances, cultural references, and colloquialisms.

\subsection{Knowledge of the contexts where generative AI is used}

Given the apparent universality of artificially generated content, users should be able to evaluate which are the appropriate contexts to use it, in which modalities, and to which extent. This requires a level of adaptation to the environmental expectations and requirements with respect to the tools and content to be used.

This competency is related to the next two, as both the legal framework and ethical values and principles influence the context where generative AI can be used. Nonetheless, this competency differs from them as it relates more closely to the specific applications of generative AI, which may vary from case to case; moreover, while the ethical and legal contexts are important, the user should also consider the social and professional context, which may determine to what extent it is appropriate and useful to use generative AI for given scenarios. 

Generative AI had a disruptive impact on different domains, and not all authorities and organizations have had the time to adapt to it by developing guidelines and rules, as in the case of the scientific and academic world \cite{stokel2023, stokel2023chatgpt}. While this is likely to change in the near future, users should in any case be able to use their discretion to adapt their use of generative AI to different circumstances.  

\textbf{Example:} Several universities have developed guidelines for the use of LLMs in assignments \cite{Cotton2023}. This differs from campus to campus, with some of them prohibiting LLMs altogether, and others allowing them for certain types of coursework. Students should learn these guidelines, and behave accordingly. 

\subsection{Knowledge of the Ethical Implications}

It is important to critically analyze both generative AI models and their outputs from a human perspective, considering how they interact with the world and society. In other words, this competency is about construing generative AI as a human artifact, and how it operates from a societal point of view. 
With this respect, users should learn how to align their use of generative AI with their ethical values, understanding the possible ethical implications both of the models and their outputs. 

Ethical aspects should be taken into consideration in all stages of the use of generative AI: when programming and deploying it, when designing guidelines to regulate its use, and when using it and employing it for specific use-cases \cite{Liebrenz2023}.

\textbf{Example:} A campaign manager showing an AI-generated image without making it explicit may induce voters to think that they are seeing a true photo rather than a fabricated one. This may mislead the public during an election, and while it may not necessarily be illegal, its ethical implications should be analyzed and considered. An example of such a scenario occurred during the campaign for Switzerland's 2023 elections \cite{Wendling2023}.

\subsection{Knowledge of Legal Aspects}

While at the time of writing this paper, only a few countries have begun a legislative process regulating AI and its applications, it is likely that several others will follow \cite{Hacker2023, Mesko2023}. Indeed, in March 2024 the European Parliament approved the European AI Act \cite{EUAIACT2021}, determining a set of norms to be followed with respect to several aspects of AI at large \cite{Helberger2023, Heidelberger2023aiact, madiega2021}. At the same time, several states from the United States have begun to adopt laws targeting specific aspects of AI \cite{chae2020us}, and in 2023 President Joe Biden issued Executive Order 14110 on Safe, Secure, and Trustworthy Development and Use of Artificial Intelligence \cite{harris2023}.

Users should be generally aware of the legal framework concerning AI in general and generative AI specifically, to know to what extent they can legally use generative AI, as well as to understand their rights and the guarantees that they have by law.

\textbf{Example:} The most prominent example of AI legislation to date is the European AI Act \cite{EUAIACT2021}, and European users should become familiar with it and its implications.

\subsection{Ability to continuously learn}

While this ability is not limited to generative AI and can be extended to other domains, the high dynamism of the field of generative AI imposes the necessity for users to be able to constantly learn new tools, functionalities, issues, and regulations that may arise. This is a transversal competency, that applies to each of the other ones, which should be regularly updated. While this skill may appear to be trivial, it is important to make it explicit so as to highlight the agency of the individual. 

\textbf{Example:} A user of video generative AI should keep up with the updates of the tools that they use, and be aware in case new tools become available.

\section{Discussion} \label{sec:discussion}

\subsection{Advantages and Limitations}

The presented competencies for generative AI literacy gather prominence with the increasing pervasiveness of generative AI technologies, exerting a deep impact on both individuals and society at large. This framework will not only equip individuals to engage with generative AI across various domains but also encourage responsible and ethical usage to mitigate potential risks. The competencies guide individuals to move from being consumers to interpreters of AI models, allowing them to optimize tool utilization for specific problem-solving objectives, and to traverse the complexities of AI-generated content with understanding and insight, thus promoting AI as a collaborator. These competencies collectively contribute to a versatile skill set, empowering individuals to actively participate in the development, application, and ethical use of generative AI across a spectrum of applications and industries. In addition, the competencies can serve as a foundation for designing educational programs and curricula focused on developing generative AI literacy depending on one's interaction level with generative AI, ranging from consumers to developers.

In the context of generative AI literacy competencies, it is paramount to recognize the complexities and challenges that underlie the development and application of such a framework. Generative AI technology is evolving rapidly, with new models and techniques continually emerging, each with its own set of intricacies. The competencies presented in this framework are designed based on the current state of the field. However, as technology advances, revisions to the competencies may be necessary.

\subsection{Relevance for the Public Sector}
 
Policymakers and regulatory bodies can employ these competencies to inform discussions and establish clear guidelines for generative AI literacy. By aligning with these competencies, policymakers can ensure that AI literacy is integrated into various policies and frameworks, promoting responsible and ethical use of generative AI technologies.

AI is already employed at various levels in the public sector, including service delivery, management, and monitoring \cite{Martijn2021, Henman2020, Geyer2019, Misuraca2022}. As part of organizations, individuals already use (or can use) generative AI applications for tasks such as note-taking, summarization, time management, and providing faster support to common queries \cite{bright2024}. In addition to its role in supporting policy development and regulation, generative AI has the potential to ease access to internal knowledge and databases through chat and also provide answers to frequently asked questions by citizens \cite{Margetts2023}. 

Understanding how generative AI will be employed by government personnel is essential for comprehensively addressing the implications of AI adoption in the public sector and shaping future governance models. Apart from being users of these tools, policymakers and bureaucrats will have to deal with various issues related to the tools, their deployment and usage, and the results they produce in terms of ethical, legal, and political implications; at the same time, accountability and transparency measures need to be put in place to safeguard against potential risks and biases. All these issues underscore the importance of equipping policymakers and government officials with the necessary knowledge and skills, built and assessed at different levels, to effectively use generative AI.

In our opinion, the proposed generative AI competency model can be used by public institutions as a starting point towards assessing the competencies of their employees, and to make institutions themselves more aware of where they stand with respect to the adoption of gen AI tools. The proposed model can also serve as a starting point to develop training material to allow public organizations to update their employees’ skills, and reformulate their strategies as the field of generative AI continues to evolve.

\subsection{Propositions for Further Research and Applications}

As the field of generative AI literacy continues to evolve, the following propositions highlight potential research directions based on identified gaps and emerging trends. Additionally, the competencies identified open up applications across various domains, some of which are discussed below.

\subsubsection{Specialized Competencies}

The framework proposed in this work provides competencies for a broader generative AI literacy. However, generative AI applications span various domains, and one direction involves the investigation of domain-specific competencies. As different industries or applications may require specialized skills and ethical considerations, finer and specific sub-skills might be needed.

\subsubsection{Assessments}
 
The competency-based model for generative AI literacy provides a footing for developing assessments that involve defining tasks and associated metrics that can measure an individual's proficiency in generative AI. These assessments can be utilized in educational settings, certification programs, or workplaces to objectively and subjectively evaluate an individual's understanding of generative AI concepts, capabilities, and ethical considerations. These measures make explicit the abilities of an individual and allow for the classification of users into different proficiency levels and the creation of teaching resources to guide educators.

\subsubsection{Curriculum Development in Educational Institutions}
 
Academic institutions can leverage these competencies as a comprehensive framework to develop AI literacy curricula catering to various levels of education, from elementary to higher education. These curricula can effectively equip students with the necessary knowledge, skills, and understanding to navigate the rapidly evolving domain of generative AI.
 
\subsubsection{Professional Development Programs}
 
Organizations and training institutions can leverage these competencies to design professional development programs. These programs can cater to individuals seeking to enhance their skills in AI-related fields or transitioning into roles involving generative AI. Such programs can also ensure that employees are equipped with the necessary skills to navigate and leverage generative AI technologies in the workplace.
 
\subsubsection{Community and Outreach Programs}
 
Non-profit organizations and community initiatives focused on AI education can utilize these competencies as a guiding framework for designing literacy outreach programs. By aligning with these competencies, these programs can effectively reach individuals from diverse backgrounds and empower them with essential AI literacy skills, fostering a more inclusive and knowledgeable society equipped to navigate the AI era.

\subsubsection{Impact Analysis}
Further research could involve assessing the long-term impact of generative AI literacy on decision-making processes. It should investigate how individuals with generative AI literacy competencies navigate ethical and practical challenges over an extended period with other people, generative AI systems, and in hybrid settings.

\subsubsection{Human - AI Collaboration}
Lastly, there is a need to create frameworks that facilitate effective collaboration between humans and generative AI systems, focusing on how competencies can be extended to enhance communication, trust, and symbiotic relationships between users and AI models for a productive collaboration.

\section{Conclusion} \label{sec:conclusion}
The set of competencies enumerated in this paper through a Competency-Based Model aims to guide research toward defining an understanding of generative AI literacy. The work lists twelve key competencies to build the knowledge and skills that are essential for understanding and using generative AI systems. The ability to comprehend, assess, and employ generative AI technologies will be fundamental to empower individuals to engage with and contribute to an AI-driven world effectively. Generative AI literacy goes beyond technical proficiency and requires combining it with other proficiencies such as adaptive usage, ethical considerations, and legal awareness. Thus, the competencies outlined here are not confined to specialists; rather, they cater to a broader audience, reflecting the growing integration of AI into various aspects of daily life. 

Developing generative AI literacy facilitates informed decision-making and participation in AI-driven initiatives and skills to understand, create, and use generative AI systems will ensure realistic expectations, enabling users to leverage AI technologies effectively while avoiding overreliance on their capabilities. By developing these capacities, individuals can use AI as a collaborator within their domain and AI as a tutor/guide for other subjects. 

Generative AI literacy is not a static set of skills, but an evolving one. The framework presented here serves as a starting point and, as AI technologies evolve, researchers should revise these competencies and update them. With growing AI applications, the integration of these competencies into educational curricula, training programs, and professional development initiatives is imperative. Stakeholders are urged to collaboratively design and implement initiatives fostering generative AI literacy at various levels. The competencies lay the groundwork for a potential global standardization of generative AI literacy and nurture a new generation of AI-literate individuals who are not passive consumers but active contributors to the ongoing developments of generative AI.

\begin{acks}
This work was supported by the European Union’s Horizon 2020 IcARUS project (under grant agreement No. 882749) and the Horizon Europe ELIAS project (under grant agreement No. 101120237). 
\end{acks}

\bibliographystyle{ACM-Reference-Format}
\bibliography{bibliography}

%\appendix

\end{document}